\documentclass[prl,twocolumn,amssymb,showpacs,floatfix]{revtex4}

\usepackage{graphicx}
\begin{document} 


\title{Magnetic Field Effects on the Transport Properties of 
One-sided Rough Wires}

\author{A. Garc\'{\i}a-Mart\'{\i}n$^*$, M. Governale$^{\dagger}$ and
P. W\"olfle$^*$} 
 
\affiliation{$^*$ Institut f\"ur Theorie der Kondensierten Materie, 
Universit\"at Karlsruhe, P.O. Box 6980, D-76128 Karlsruhe, Germany\\
$^{\dagger}$ Institut f\"ur Theoretische Festk\"orperphysik,
Universit\"at Karlsruhe, D-76128 Karlsruhe, Germany} 

\date{\today} 
 
\begin{abstract}    

We present a detailed numerical analysis of the effect of a magnetic
field on the transport properties of a `small-$N$' one-sided 
surface disordered wire. 
When time reversal symmetry is broken due to a  magnetic field $B$,   
we find a strong increase with $B$ not only of the localization  
length $\xi$ but also of the mean free path $\ell$ caused by boundary states. 
Despite this, the universal relationship between $\ell$  and $\xi$ does
hold. We also analyze the conductance distribution at the metal-insulator
crossover, finding a very good agreement with Random Matrix Theory 
with two fluctuating channels within the Circular
Orthogonal(Unitary) Ensemble in absence(presence) of $B$.
\end{abstract}  
\pacs{72.15.Rn, 72.10.Fk, 42.25.Dd, 71.30.+h } 
\maketitle 

The physics of disordered media has attracted the attention of scientists
from different areas since long ago\cite{Reviews}, and it is an up-to-date
topic due to its relevance in wave transport, be it in the form of  
quantum (electrons) or classical (electromagnetic or acustic) waves.
Transport in disordered media, in particular in the 
mesoscopic regime, presents a rich panoply of phenomena, such as  
\emph{Anderson localization}\cite{Anderson},
\emph{Weak localization}\cite{Weak}, and
\emph{Universal Conductance Fluctuations}\cite{UCF}, just to mention a few. 

When dealing with disordered systems, 
it is of fundamental importance to
know about the full statistical properties of the relevant transport
quantities, i.e. the conductance $G$, the eigenchannels $\tau$ and the
transmission  $T_{i},T_{ij}$ and reflection $R_{i},R_{ij}$ coefficients.
Much is known about statistics of both the transmission\cite{Langen,PRL1}
and the reflection\cite{PRL2} coefficients, since the full distribution
function has been obtained from a Random Matrix Theory (RMT) 
approach\cite{Langen,PRL2} and numerical techniques\cite{PRL1,PRL2} as well as
experimentally in the microwave regime\cite{Stoytchev}.  

Several approaches have been considered to analyze the effect of disorder
on the transport properties, ranging from analytical techniques 
using RMT\cite{DMPK,RMT,Pichard,muttalib} to numerical 
ones\cite{Numerical,gatoprl}. Most of them consider that the disorder is
inside the material where the wave travels through (bulk disorder). 
However, the recent
developments in nanotechnology open the way to new
sources of scattering. In particular, as system sizes are
shrunk down to the nanometer scale,  the surface-to-volume ratio becomes
larger, and surface effects (for example disorder) 
become very important.   This
is more relevant in the case of  few-modes quantum wires,   where the
effect of surface disorder (roughness) can dominate their  transport
properties.  In addtion, due to improvements in fabrication  techniques
bulk disorder is almost absent. Many statistical properties of these
surface disordered wires  have been shown to be in very good agreement
with those predicted by RMT\cite{APL,PRL1,PRL2,surface,gatoprl}.

The consequences of time reversal symmetry breaking in disordered media,
such as the doubling of the localization length as a consequence of the
phase mismatch of time reversed paths, are conceptually known since long ago
(see for instance \cite{Reviews,RMT}). Nevertheless, a fundamental question
was still not answered: how does the transition take place? This issue has
recently been addressed, both theoretically\cite{kolesnikov} and
numerically, by means of an  Anderson model (bulk
disorder)\cite{schomerus}, finding that the  expected doubling of the
localization length takes place in a rather smooth way, while the mean
free path remains constant.  

Another conceptual issue concerns the shape of the full conductance distribution
$P(G)$ in the diffusion-localization transition. It is known that $P(G)$
evolves from Gaussian, in the diffusive regime, to log-normal in the
localization regime. It has been  recently  predicted\cite{muttalib}
that,  in the crossover regime, $P(G)$ exhibits a highly non trivial shape; 
the subsequent numerical analysis revealed a direct relation between the
number of eigenchannels that actually describe the distribution and its
universal characteristics\cite{gatoprl}.

\begin{figure}
\includegraphics[width=8cm]{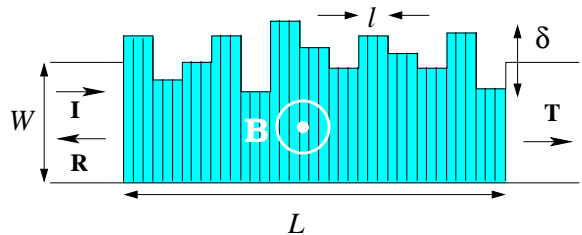}  
\caption{Schematic view of a nanowire with one-side surface disorder.} 
\label{System}  
\end{figure}

In this paper we analyze the consequences that the application of a
magnetic field has on the transport properties of a one-sided 
surface disordered wire (see Fig.~\ref{System}). We
show that breaking time reversal symmetry leads to drastic effects on   
the electron propagation
through rough wires in the small-number-of-channels regime.
The dependence  of the mean free path $\ell $ and localization length $\xi$ on 
the magnetic field strength is studied by  means of an analysis of
$\langle 1/G\rangle$ and $\langle \ln G\rangle$ as a function of both the
length and the magnetic field strength. 
This analysis reveals a peculiar behavior of this two relevant
length scales: $\xi$ increases, but much more than the expected factor of
two\cite{kolesnikov,schomerus}, while at the same time $\ell$, which is 
expected to remain constant (for the low fields considered),  
also increases.
We find that the latter phenomenon is  related to the precursors of the 
edge states (we are far away from the quantum Hall regime), which 
propagate almost freely due to the specific type of disorder. We show
that the unexpected large increase of $\xi$ does 
not contradict the predictions of RMT
when the anomalous behavior of $\ell$ is taken into account.  The validity
of RMT is confirmed by the analysis of the numerical conductance
distribution $P(G)$ in the vicinity of the diffusion-localization
transition. This analysis shows that the universal features of the
distribution are not affected by the lifting of time reversal invariance.

The system which we are dealing with (depicted schematically in 
Fig.~\ref{System}) is  made of two semi-infinite two-dimensional clean 
leads of width $W$ and a surface-disordered region of length $L$ in between.  
The corrugated region is composed of slices of length $\ell$ laterally bounded 
by hard walls, whose random 
width is uniformly distributed in the interval $[W-\delta,W+\delta]$ around 
the mean value $W$. We will take $W/\lambda=2.6$ ($\lambda$ being the
electron Fermi wavelength), which allows  
five propagating channels in the
clean part $N=5$, and a disorder strength $\delta$ so that 
$W/\delta=7$\cite{note0}. 
We consider 
\emph{One-sided Surface Disorder} (OSD), i.e., the roughness is located 
only on one surface. 
In order to study how the presence or absence of time reversal  symmetry
affects the transport, a magnetic field $B$ perpendicular to the plane of the
wire  is considered. 

The conductance (in units of $2e^2/h$) of the total
system is simply given by  $G={\mathrm{trace}}\{tt^\dagger\}$.
The transmission matrix $t$ is calculated for each realization of the 
disorder by a generalized scattering matrix 
technique\cite{APL,WeissJosanMetodo} where the magnetic 
field has been introduced in the formalism as in the method of Ref.\cite{gvrboese}. 
This consists in choosing a gauge, in which only the component of the vector 
potential along the transverse direction is non zero; the system is then 
divided in discretization slices (separated by thin lines in Fig.~\ref{System}), 
and a constant vector potential is assigned to each slice. To minimize 
errors due to the discretization, the slices have to be chosen so 
that they contain less than a 
flux quantum, furthermore convergence can be checked by 
increasing the number of them.
Then, in order to obtain the statistical features introduced by the
disorder, we perform configuration averages $\langle \cdots \rangle$. 
To calculate the distributions we average  over 10000 different
configurations of the disorder. 

\begin{figure}
\includegraphics[width=8cm]{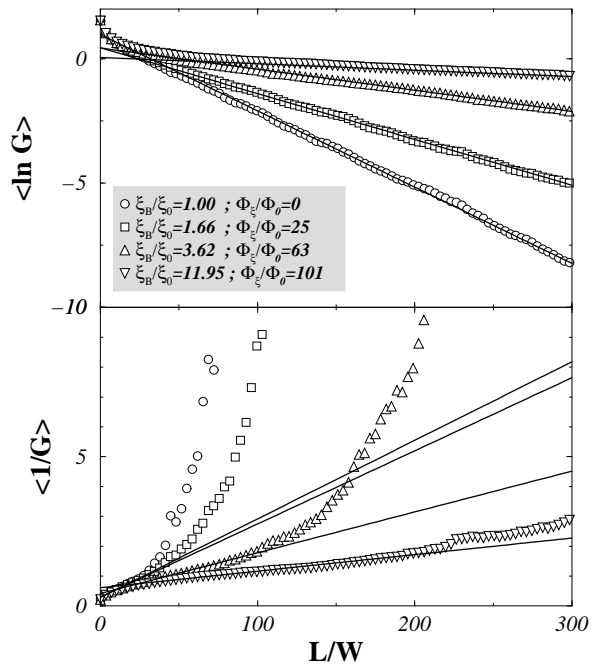}   
\caption{ 
Logarithm of the conductance (upper panel) and Resistance (lower panel) as 
a function of the length of the disordered
region  for different values of the magnetic field (symbols). 
The lines indicates the best linear fits (obtained in the 
localization regime for the logarithm of the conductance, and in the 
diffusive regime for the resistance), 
which are used to extract respectively the localization length, 
and the mean free path.} 
\label{logG-R} 
\end{figure}

In the upper panel of Fig~\ref{logG-R} we plot $\langle \ln G \rangle$
versus $L/W$ for four different values $B$. The magnetic field strength is 
given by the 
quantity $\Phi_\xi=\xi_0WB$ that measures the amount of magnetic flux
trapped in a portion of wire with a length equal to the localization length
at zero field ($\xi_0$). Throughout this work all magnetic fluxes are given 
in units of the flux quantum $\Phi_0$. 

 The localization length $\xi_B$ for the different values
of $B$  is obtained from the linear part of the plot since 
$\langle\ln G\rangle\propto -L/\xi_B$. The value of $\xi_B$ evolves
as expected, the higher $B$ the larger $\xi_B$, but instead of  
saturating to the expected value 
$\xi_B=2\xi_0$\cite{RMT,Pichard,kolesnikov,schomerus} when
time reversal invariance is broken, for OSD 
this does not occur giving rise to values $\xi_B\gg2\xi_0$.
 
The origin of this lies in the fact that, whereas usually $\ell$ is 
assumed to be  
constant for the small values of $B$  that are considered (see, for
example, the results  for a two-dimensional Anderson model with on-site
disorder of  Ref.~\cite{schomerus}), in wires with OSD one has to take the
magnetic field dependence of $\ell$ into account.   This effect
can be seen in  lower panel of  Fig.~\ref{logG-R}, where the average
resistance $\langle1/G\rangle$ is  plotted as a function of $L/W$. By virtue of 
$\langle 1/G\rangle \approx 1/N +  L/(N \ell)$ (as long as $\ell < L < \xi
$) the  best linear fit in  the diffusive regime gives $\ell_B$.

The general relationship between mean free path and localization length
predicted by RMT from general symmetry considerations can be written as
\begin{equation}
\xi_{\text{B}}=\beta N \ell_{\text{B}},
\label{locl} 
\end{equation}
with $\beta$ being 1 in 
the presence of time reversal symmetry, and 2 when it is 
broken  (by the magnetic field). 
In Eq.~(\ref{locl}) the magnetic field dependence of the 
mean free path has to be accounted for.

\begin{figure}   
\includegraphics[width=8cm]{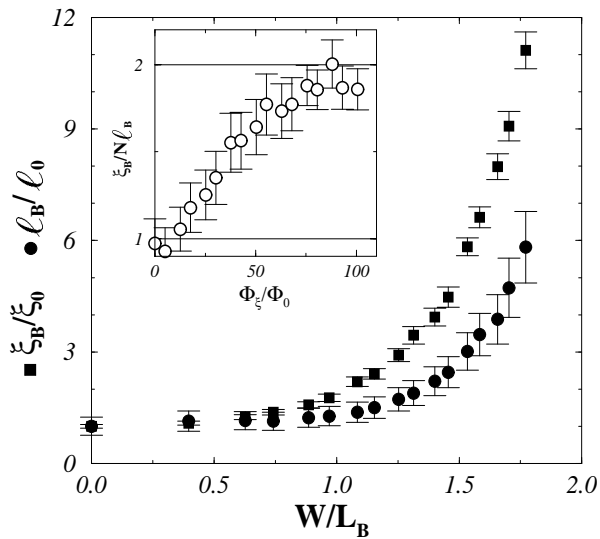}
\caption{Magnetic dependence of the mean free path and of the 
localization length.  
Both the mean free path and the localization length are normalized to their 
zero-field value;  
the magnetic field is in units of $W/L_{\text{B}}$, with $L_{\text{B}}=
\sqrt{\hbar/e B}$, being the magnetic length. 
Inset: localization length divided 
by the number of modes and the mean free path; this quantity reaches the 
value 2  for 
fully broken time reversal symmetry. 
} 
\label{xi-l} 
\end{figure}

Eq.~(\ref{locl}), predicts a transition from $\beta=1$ to $\beta=2$,  as
soon as time reversal symmetry is removed, i.e. when the magnetic flux
comprised in a section  of wire of length of the order of $\xi_0$ is about
one flux  quantum.  As shown in  Fig.~\ref{xi-l}, both $\xi_B$ and
$\ell_B$ increase smoothly when $B$ increases and in the same qualitative
way.  In fact, the tremendous increase of $\ell_B$, allows the
possibility  to have  huge  values of $\xi_B$   without violating
Eq.~(\ref{locl}) in any case (see inset in Fig.~\ref{xi-l}). The huge
increase of these two quantities gives rise to the actual possibility of
changing the length of the diffusive regime with a 
magnetic field. The reason for that lays in the presence of one 
clean surface. In fact, even for really small $B$, the developing
 edge states may 
propagate along the clean surface suffering almost no scattering. 
We have checked that this is indeed the case 
by doing the same analysis for a wire with 
\emph{two-sided surface disorder}, with 
uncorrelated disorder of 
equal strength on both surfaces (not shown here). 
In the absence of the clean surface, 
the non fully developed edge states still suffer from scattering, and no 
appreciable variation of the mean free path is found. 
It is important to stress once more 
that the  magnetic fields that give rise to the aforementioned effect 
are small (cyclotron radius much bigger than wire 
width). Thus, we are \emph{far away} from the 
quantum Hall regime, where it is natural that edge states 
propagate freely without suffering any scattering.   
We note in passing 
that the magnetoresistance of these one-sided rough wires may offer 
interesting technical possibilities.
 
\begin{figure}
\includegraphics[width=8cm]{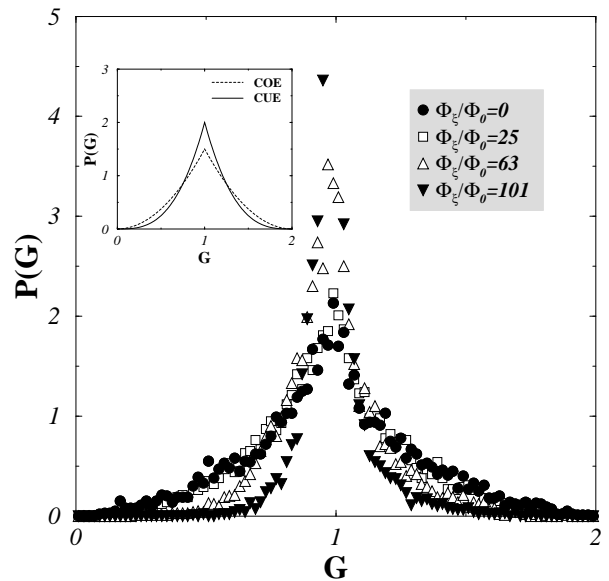}  
\caption{ 
Conductance distribution close to the diffusion-localization transition for
different strength of the magnetic field. In the inset the
results from RMT for the maximum entropy Ansatz are shown. 
The distributions exhibit an increase of the cusp height as well as a
decrease of their width with increasing magnetic field.} 
\label{P_G} 
\end{figure}

It is worth to mention that 
our numerical simulations show that the lifting of time
reversal invariance is a gradual process, with $\beta=2$
reached for around 100 flux quanta per localization length times system 
width, as it is 
shown in the inset of Fig.~\ref{xi-l}. This result is in perfect agreement with 
the values presented in Ref.~\cite{schomerus} even though several
characteristics of the transport depend on the kind of disorder. 
We also see no evidence of the 'two-scale localization' predicted in
Ref.~\cite{kolesnikov} and not found in Ref.~\cite{schomerus}.
 
Now we go beyond the first moments of the distribution, and devote 
our attention
to the entire distribution function $P(G)$ at  the transition point. It has
been recently reported\cite{gatoprl} that $P(G)$ contains some universal
characteristics that  depend only on the value of $\langle G \rangle$, and
in particular, that the transition point corresponds to a transition in
the number of active eigenchannels. We would like to address these issues, 
when the action of the magnetic field eliminates time reversal
invariance, by analyzing $P(G)$ in the vicinity of the
diffusion-localization transition. Let us focus on the characteristic
features of $P(G)$ just before the localization onset, more explicitly when
$\langle G\rangle=1$. At this value of the mean conductance the
distribution has a well define cusp at $G=1$. 
In  Fig.~\ref{P_G} we show that $P(G)$ for $\langle G\rangle=1$ shrinks
with increasing $B$ and at the same time the cusp becomes higher and more
pronounced.  
This confirms that general characteristics  of $P(G)$ at the
transition point actually does not depend on the time reversal invariance
conditions.

It has been shown
that the RMT picture\cite{Baran94} within the maximum entropy
Ansatz (this leads to the Circular Orthogonal Ensemble (COE) when   
time reversal symmetry is present \and to the Circular Unitary 
Ensemble (CUE) when
it is broken) gives an almost quantitative description of the
distributions when time reversal symmetry is preserved\cite{gatoprl}.
Following the ideas of Ref. \cite{gatoprl} we obtain $P(G)$ for two active
eigenchannels (Eq.~\ref{RMT_n2}):

\begin{eqnarray}
P(G,\langle G\rangle)\propto&& \frac23 \left(1-\frac G2\right)^3
\left(\frac G2\right)^{2\alpha}\times\nonumber\\
& & {}_2F_1\left(\frac32,-\alpha,\frac52,\left(\frac 2G -1\right)^2\right)
\Theta(G-1) \nonumber\\&+& \frac{\sqrt{\pi}}{2}\left(\frac G2\right)^{2\alpha+3}
\frac{\Gamma (1+\alpha)}{\Gamma (\frac52+\alpha)} \Theta(1-G) \label{RMT_n2}.
\end{eqnarray}
where 
\begin{equation}
\alpha=\frac{4(\langle G\rangle-1)}{2-\langle G\rangle}\ ,
\end{equation}
depends only on the value of the mean conductance\cite{note1}.

This distribution is shown in in the insets of Fig.~\ref{P_G}, 
together with Eq. 3 from Ref. \cite{gatoprl} for COE. The qualitative 
agreement with the numerical data is striking.   

In conclusion, we have presented extensive numerical simulations for  
nanowires with one-sided surface disorder in the presence of  magnetic fields. 
The transition between the orthogonal and unitary case has been 
demonstrated to happen smoothly. 
Although  our model of disorder exhibits a large increase of 
the mean free path leading to giant wave delocalization, the 
relationship between mean free path and localization length predicted by 
RMT is fulfilled. 
The behavior of both mean free path and localization length suggest the
possibility of  manipulating the diffusion and localization onsets by 
applying a magnetic field. 
The conductance distribution around the metal-insulator transition has 
confirmed its universal nature, being very well described 
by RMT with the maximum entropy Ansatz and two fluctuating channels.   
  
We are gratefully indebted to F. Evers, A.D. Mirlin, J.J. S\'{a}enz and A. Rosch
for valuable discussions.  AGM acknowledges support from the
Emmy-Noether program of the DFG under Grant No. Bu 1107/2-1. This work 
was partially supported by the DFG Center for Functional Nanostructures at 
the University of Karlsruhe.


\begin{references}
\bibitem{note0} We have verified that changing the values of the
  parameters, in the small-N,  weak scattering regime, does not affect the
  physical  picture described in this work.
\bibitem{note1} This expression is valid for $2/3<\langle
G\rangle\lesssim 5/4$, 
i.e. when the distribution can be described by two fluctuating channels.
\end{references}
\end{document}